# The Wide-Field X and Gamma-Ray Telescope ECLAIRs aboard the Gamma-Ray Burst Multi-Wavelength Space Mission SVOM

P. Mandrou[a], S. Schanne[b], B. Cordier[b],
R. Pons[a], D. Barret[a], C. Amoros[a], K. Lacombe[a], M. Fesquet[b], O. Limousin[b],
P. Sizun[b], F. Lebrun[b,c], F. Gonzalez[d] and M. Jouret[d]

[a] *CESR/CNRS/UPS, 9 Avenue Colonel Roche, BP 44346, 31028 Toulouse, Cedex 4, France*
[b] *CEA Saclay, DSM/IRFU (former DAPNIA), 91191 Gif sur Yvette Cedex, France*
[c] *APC, 10 Rue Alice Domont et Léonie Duquet, 75205 Paris Cedex 15, France*
[d] *CNES, CST, 18 Avenue Edouard Belin, 31401 Toulouse Cedex 9, France*

**Abstract.** The X and Gamma-ray telescope ECLAIRs is foreseen to be launched on a low Earth orbit (h=630 km, i=30°) aboard the SVOM satellite (Space-based multi-band astronomical Variable Objects Monitor), a French-Chinese mission with Italian contribution. Observations are expected to start in 2013 (Wei, J., these proceedings). It has been designed to detect and localize Gamma-Ray Bursts (GRBs) or persistent sources of the sky, thanks to its wide field of view (≈2 sr) and its remarkable sensitivity in the 4-250 keV energy range, with enhanced imaging sensitivity in the 4-70 keV energy band. These characteristics are well suited to detect highly redshifted GRBs, and consequently to provide fast and accurate triggers to other onboard or ground-based instruments able to follow-up the detected events in a very short time from the optical wavelength bands up to the few MeV Gamma-Ray domain.

## INTRODUCTION

The detection and fine localization (better than 10 arc minutes) of X and Gamma-Ray Bursts (GRBs) in the late 1990's by BeppoSAX first [1], and more recently HETE II [2] and INTEGRAL (IBIS) [3], was at the origin of the discoveries of the corresponding counterparts thanks to follow-up observations by ground-based telescopes at different wavelengths (optical [4], radio [5]). They have revealed that gamma-ray bursts were associated to peculiar supernovae. We know now since then that these events occur at cosmological distances, with redshifts measured today, in the light of the high quality results recently provided by Swift [6], in a range from $z \approx 0.033$ up to $z > 6$.

In the next years, GRB studies will undoubtedly shed a new light on the evolution of the young Universe if they are accurate and complete enough to allow a good understanding of the phenomena. In order to fulfill this exciting task, future GRB studies must rely on the availability of a continuous flow of accurate GRB positions and acquisition of many additional parameters (e.g. redshift, $E_{peak}$, jet-break time, etc…) obtained through both prompt data acquisition by on-board instruments and near real-time observations by large ground-based instruments.

The next scheduled missions will have to take into consideration these requirements to increase the expected scientific return from the analysis of those objects and consequently their understanding. It has been demonstrated by the Swift mission that the detection of highly redshifted GRBs was possible; a sample of 7 GRBs has already been detected with $z > 4$. An exceptionally high-redshift event, GRB 050904 [7] with $z \approx 6.29$, has been discovered, exhibiting after refined analysis a weak and particularly long tailed emission in the BAT detector ($T_{90} > 170$ s), and an $E_{peak}$ in the energy range below 30 keV (where $T_{90}$ is the time interval over which 90% of the total background-subtracted counts are observed, with the interval starting when 5% of the total counts have been observed). The largest instruments available by now in space, such as Swift [8] or INTEGRAL-IBIS [3], are sensitive above 15 keV.

This means clearly that they will be facing difficulties to detect event at large redshifts, due to the $E_{peak}$ approaching their low-energy threshold. HETE II had the potential to detect those events, but its triggering detector lacked sensitivity, let alone the fact that it didn't have imaging capabilities needed to detect weak and long duration events.

Thanks to new technological developments, we have designed a new instrument, called ECLAIRs, able to detect and trigger on gamma-ray bursts at lower energies, in the 4-70 keV energy range.

The technical design requirements imposed on this instrument by the scientific needs are:
- allow the detection of all known types of GRBs ($\approx$ 80 GRB triggers per year)
- provide fast reliable GRB positions ($\Delta\theta$ < 10 arc min.) within $\Delta t \approx$ few seconds
- deliver fast triggers and GRB parameters to onboard and ground-based instruments ($\Delta t$ < 1 min)
- measure in real time the specific parameters of the prompt emission (light curve, spectra).

## THE ECLAIRS INSTRUMENT CONCEPT AND PERFORMANCES

The ECLAIRs instrument [9] is made of a detector plane mosaic built with 6400 CdTe detector pixels (4 mm × 4 mm × 1 mm thick, 1024 cm$^2$ active area), distributed as a 80×80 chess-board pattern arrangement (4.5 mm pitch, 36×36 cm side length), assembled from 200 elementary monolithic structures (XRDPIX) of 32 detectors (8×4), analyzed each by a 32-inputs analog front-end electronics ASICS. An electronic box (UGD) is in charge of all the detector plane management; it provides the needed voltages to power the ASICS and their analyzing electronics, the reverse high-voltage bias to polarize the detectors (-600 V), and all the numeric contexts to set-up the working parameters of the 6400 independent detectors managed by a FPGA processor. To reach the low-energy threshold of 4 keV, these detectors are cooled down to -20 °C thanks to a cold plate powered by a semi-active cooling system built around a large area radiator and variable-conductance heat pipes. A PID software, processed in real-time by the FPGA, is in charge of the regulation and stabilization of the detectors working temperature. The detection plane data are analyzed by 8 analog to digital converter assemblies (ELS), each of which is able to encode the analog signals of 25 XRDPIX and to send the coded energy (1024 channels in the range 4-250 keV) to the digital processing electronics box, called UTS [10]. The absolute detection time of each event is also measured and sent to the UTS with an accuracy of 10 μs, as well as the (X,Y) position of the hit detector pixel.

A coded aperture mask, made of a 0.6 mm thick Tantalum sheet, placed 46 cm above this detection plane, defines a coded field of view of $\approx 90°\times90°$ ($\Omega$ > 2 steradians).

To reduce the contribution of the diffuse background component in the low energy range, hence to increase the telescope sensitivity, a graded shield made of Lead (1 mm thick), Copper (0.1 mm) and Aluminum (0.5 mm) is placed around the detector up to the mask, in order to limit the field of view of the detectors to the one of the coded aperture mask.

The imaging capabilities, available up to $\approx$ 120 keV, allow a localization accuracy of sources better than 10 arc minutes for a 7 σ confidence level detection.

Data are continuously analyzed on board by the scientific processing and triggering unit (UTS), in order to detect burst events, either by first detecting a count-rate increase (in several energy bands) on various time scales from 10 ms to 20 s followed by the construction from the triggered time window of an image in which a new source is searched, or by systematic searches for new sources in images built on time scales longer than 20 s.

The on-board detection and localization of a gamma-ray burst events (with an accuracy $\Delta\theta$ < 10 arcmin), delivers a general alert trigger sent to ground over a dedicated VHF network covering the orbit, in a very short time ($\tau$ < 60 s) to be used by ground-based instruments.

The trigger is simultaneously sent to the service module of the SVOM satellite [11] to initiate a slew maneuver (when safely possible regarding the pointing constraints) in order to point accurately the optical axis of the on-board co-aligned narrow-field-of-view instruments, operating at different wavelengths, in the direction of the localized event for a fast follow-up of the event and position refinement.

The in-flight performance has been calculated for the scheduled flight orbit, using a geometrically representative model of the detector and a GEANT-4 simulation of particle interactions in its materials [12]. The photon and particle fluxes, as well as the spectral parameters used to deduce the background, have been taken from the normalized data basis available for such an orbit.

The major components of the ECLAIRs instrument are presented on Figure 1; the main scientific characteristics resulting from the design described above are summarized in Table 1.

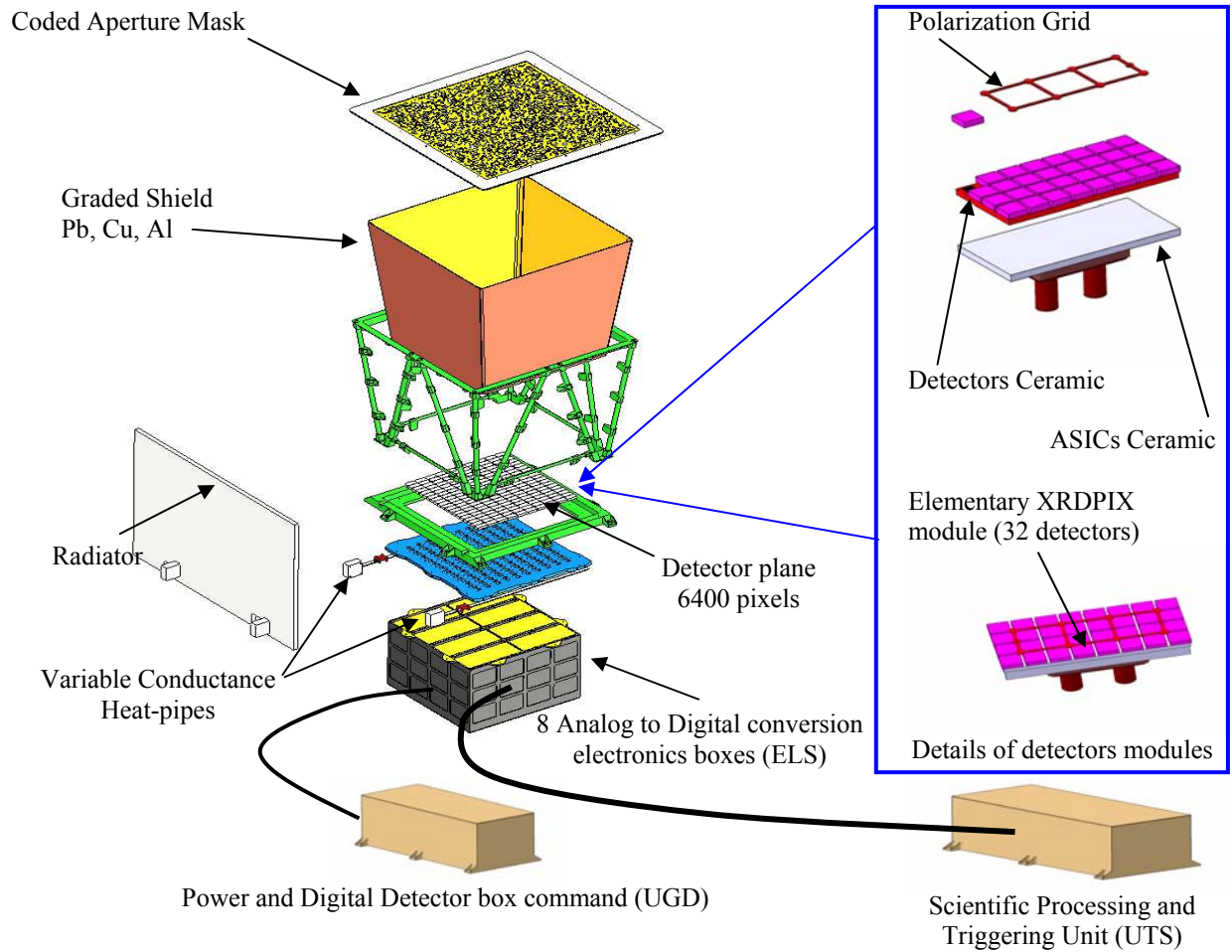

**FIGURE 1.** Components of the ECLAIRs Telescope

| Low Energy Threshold | 4 keV |
|---|---|
| High Energy Threshold | 250 keV |
| Enhanced Imaging sensitivity | 4 keV to ≈ 80 keV |
| GRB duration detection range | from 10 ms to 1000 s |
| Field of view | 2 sr |
| Detector sensitive area | 1024 cm$^2$ (80 × 80 pixels) |
| Photon timing resolution | 10 μs |
| Energy resolution | < 2 keV FWHM at 60 keV |
| Expected GRB detection and localization rate | ≥ 80 per year with > 7 σ level detection |
| Source localization accuracy | < 10 arcmin with > 7 σ level detection |
| Satellite slew maneuver duration | Δt ≈ 1 min after trigger (slew speed ≈ 10°/min). |

**TABLE 1.** Characteristics of the ECLAIRs Telescope

On the following figures are summarized the estimated performance for the ECLAIRs telescope, obtained from the numerical simulations. On Figure 2 are shown, on the left side, the calculated background as the result of two components (the main component, below 90 keV results from the diffuse contribution, and the second one, above 100 keV is due to particle-induced background). On the right side is reported the deduced flux sensitivity limit compared to previous instruments.

On the left side of Figure 3 is shown the localization accuracy (in arc minutes vs. the Signal to Noise Ratio in σ units) resulting from an on board image reconstruction (theoretical limit compared to simulated one). In the same figure is also shown the fraction of events localized (as a plateau varying from 0 to 1); above a 5σ detection level,

90% of the sources can be localized. On the right side are shown 333 SWIFT GRBs presented on a diagram (in photon fluence vs. $T_{90}$). The ECLAIRs (6 σ) sensitivity is reported on this view for a GRB at 30° from optical axis (dashed line) and on axis (solid line). The main part of the SWIFT GRBs would be detected by ECLAIRs including the most redshifted ones (so far all Swift GRBs with z>3 would be detected by ECLAIRs).

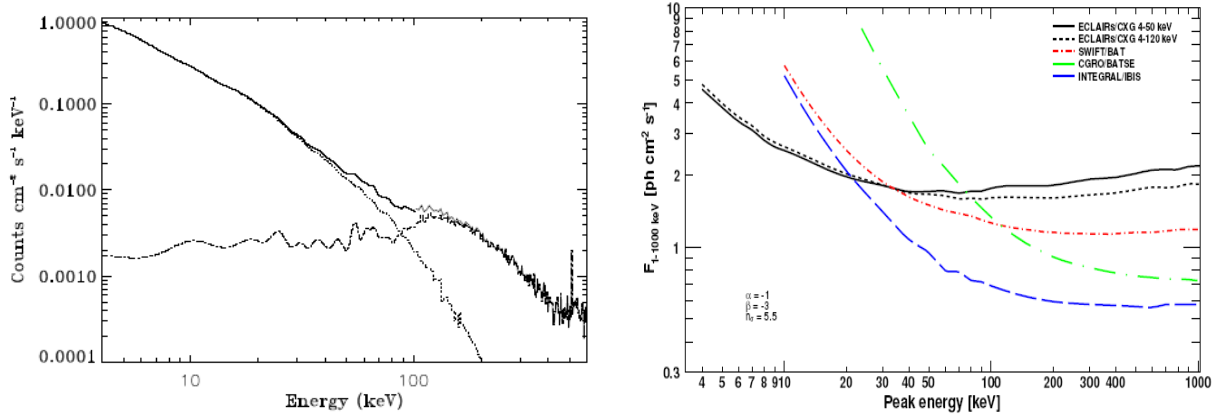

**FIGURE 2.** Background vs photon energy (left) and flux sensitivity limit as a function of the GRB peak energy (right)

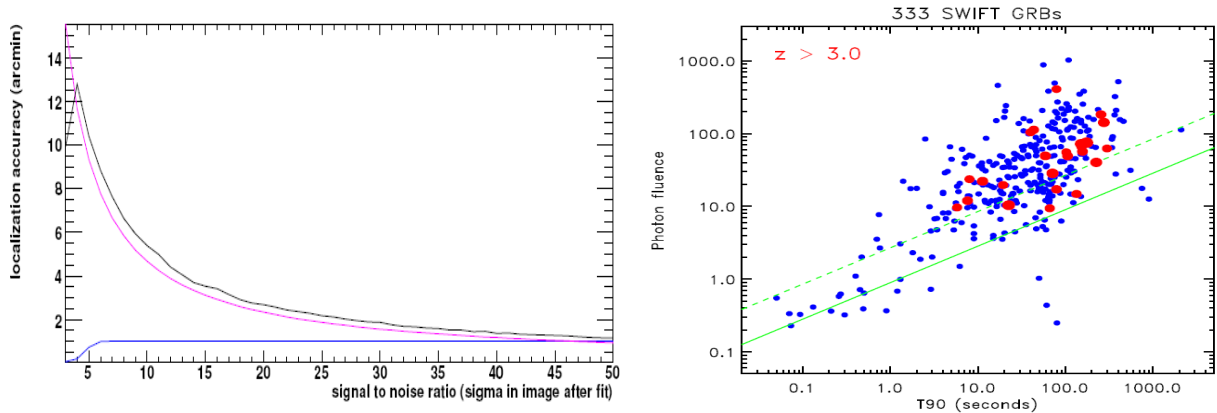

**FIGURE 3.** GRBs localization accuracy and sensitivity to redshifted events

The results presented here above demonstrate that by combining a low-energy threshold of 4 keV, a relatively large effective area and a wide field of view, ECLAIRs holds great potential for the study of highly redshifted gamma-ray bursts, thus offering a new tool to study the Universe in its youngest age.